\documentclass[12pt]{article}
\usepackage{amssymb,amsmath,amsfonts,   amsthm,nomencl,mathrsfs } 
\usepackage{bbm}
\usepackage[arrow, matrix, curve]{xy}
\usepackage[latin1]{inputenc}
\usepackage{a4wide}
\usepackage{color}

\newcommand{\IC}{\mathbb{C}}
\newcommand{\IR}{\mathbb{R}}

\newcommand{\question}[1]{\leavevmode{\marginpar{\tiny%
$\hbox to 0mm{\hspace*{-0.5mm}$\leftarrow$\hss}%
\vcenter{\vrule depth 0.1mm height 0.1mm width \the\marginparwidth}%
\hbox to 0mm{\hss$\rightarrow$\hspace*{-0.5mm}}$\\\relax\raggedright #1}}}

\newcommand{\ISS}{\mathscr{S}}

\newcommand{\IN}{\mathbb{N}}

\newcommand{\IP}{\mathbb{P}}

\newcommand{\IE}{\mathbb{E}}

\newcommand{\Id}{{d}}

\newcommand{\f}{\frac}
\newcommand{\nn}{\nonumber}

\newcommand {\rr}{\mathbb{R}}

\newcommand{\one}{\mathbbm{1}}
\newcommand{\id}{\mathbbm{1}}

\theoremstyle{plain}            
\newtheorem{theorem}{theorem}[section]

\newtheorem{Corollary}[theorem]{Corollary}
\newtheorem{Theorem}[theorem]{Theorem}

\theoremstyle{definition}       
\newtheorem{Definition}[theorem]{Definition}

\newtheorem{Remark}[theorem]{Remark}

\allowdisplaybreaks[1]

\title{H\"older estimates for magnetic Schr\"odinger semigroups in $\IR^{d}$ from mirror coupling}
\author{Oliver F\"urst\footnote{Universit\"at Bonn. E-mail: ofuerst@math.uni-bonn.de, corresponding author} and Batu G\"uneysu\footnote{Humboldt-Universit\"at zu Berlin. E-mail: gueneysu@math.hu-berlin.de} }

\begin{document}

\maketitle

\begin{abstract} We use the mirror coupling of Brownian motion to show that under a $\beta\in (0,1)$-dependent Kato type assumption\footnote{which is satisfied under a suitable $L^q$-assumption on the electro-magnetic potential, where $q$ depends on $\beta$ and the dimension $d$} on the possibly nonsmooth electro-magnetic potential, the corresponding magnetic Schr\"odinger semigroup in $\IR^d$ has a global $L^{p}$-to-$C^{0,\beta}$ H\"older smoothing property for all $p\in [1,\infty]$, in particular all eigenfunctions are uniformly $\beta$-H\"older continuous. This result shows that the eigenfunctions of the Hamilton operator of a molecule in a magnetic field are uniformly $\beta$-H\"older continuous under weak $L^q$-assumptions on the magnetic potential. 
\end{abstract}

\section{Introduction}

Kato \cite{Kato} has shown that each eigenfunction $\Psi$ of a multi-particle Schr\"odinger operator $H=-\Delta+W$ in $L^2(\IR^{3m})$ with a potential $W:\IR^{3m}\to \IR$ of the form
\begin{align*}
&W(x)=\sum_{1\leq j\leq m } w_{j}(\mathbf{x}_j)+ \sum_{1\leq j<k\leq m } w_{jk}(\mathbf{x}_j-\mathbf{x}_k),\\
&\text{with $w_j,w_{jk}\in L^{p}(\IR^{3})+L^{\infty}(\IR^{3})$ for some $p\geq 2$}
\end{align*}
is uniformly $\beta$-H\"older continuous for all $0<\beta< 2-3/p$, that is,
\begin{align*}
\sup_{x,y\in\IR^{3m}, x\neq y} \frac{|\Psi(x)-\Psi(y)|}{|x-y|^{\beta}}<\infty,
\end{align*}
where we have written points in $\IR^{3m}$ in the form $x=(\mathbf{x}_1,\dots, \mathbf{x}_m)$, where $\mathbf{x}_j\in\IR^3$ for $j=1,\dots,m$. In particular, an application of this result to multi-particle Coulomb type potentials shows that all molecular Hamilton operators (in the infinite mass limit) are uniformly $\alpha$-H\"older continuous for all $0<\alpha< 1$. Kato's proof relies on the Fourier transform and so does not apply directly to magnetic Schr\"odinger operators (even if one assumes a Coulomb gauge). The aim of this paper is to use probabilistic techniques to find a variant of Kato's regularity result that applies to a magnetic Schr\"odinger operator $H(A,V)$ with magnetic potential $ A:\IR^{d}\rightarrow\IR^{d}$ and electric potential $V:\IR^d\to\IR$. To this end we prove the following smoothing result (cf. Theorem \ref{main2}): \vspace{1mm}

Let $\beta \in (0,1)$ and let $C^{0,\beta}(\IR^d)$ denote the space of uniformly $\beta$-H\"older continuous functions on $\IR^d$, with its seminorm given by 
\begin{align}\label{asdfa}
\left\|f\right\|_{C^{0,\beta}}:=\sup_{x,y\in\IR^d, x\neq y} \frac{|f(x)-f(y)|}{|x-y|^{\beta}},
\end{align}
and consider for $q\in [1,\infty]$ the Banach space $C^{0,\beta}(\IR^d)\cap L^q(\IR^d)$ with its norm
$$
\left\|f\right\|_{C^{0,\beta}\cap L^q}:=\left\|f\right\|_{C^{0,\beta}}+\left\|f\right\|_{L^q}.
$$
\emph{Then for all Borel functions $ A:\IR^{d}\rightarrow\IR^{d}$, $V:\IR^d\to\IR$ with
\begin{align*}
\max\big(\left| A\right|^{\frac{2}{1-\beta}},  |\mathrm{div}( A)|^{\frac{1}{1-\beta}}\big)\in\mathcal{K}(\IR^d),\quad V\in\mathcal{K}^{\beta}(\IR^d),
\end{align*}
and all $t>0$, $1\leq p\leq q\leq\infty$ one has
\begin{align}\label{supero2}
e^{-t H (A,V)}: L^p(\IR^d)\longrightarrow C^{0,\beta}(\IR^d)\cap L^q(\IR^d),
\end{align}
and the norm of this operator can be estimated explicitly.}\vspace{1mm}

Above, $\mathcal{K}^{\beta}(\IR^d)$, $\beta\in [0,1]$, denotes the $\beta$-Kato class (cf. Definition \ref{kat} below) of Borel functions $\IR^d\to\IR$ which has been introduced in \cite{baturcd}, so that $\mathcal{K}(\IR^d):=\mathcal{K}^{0}(\IR^d)$ is the classical Kato class \cite{aizen} and one has $\mathcal{K}^{\beta}(\IR^d)\subset \mathcal{K}^{\alpha}(\IR^d)$ if $\beta\geq \alpha$. Note also that $ H (A,V)\Psi=\theta \Psi$ implies $e^{-t H (A,V)}\Psi=e^{-t \theta}\Psi$, so that one also obtains global $\beta$-H\"older regularity for eigenfunctions.\vspace{1mm}

The mapping property
$$
e^{-t H (A,V)}: L^p(\IR^d)\longrightarrow L^q(\IR^d)\cap C(\IR^d),
$$
is well-known \cite{BroHunLes} and only requires a local Kato assumption on $\left| A\right|^2$, $|\mathrm{div}( A)|$ and the positive part of $V$, and a global Kato assumption on the negative part of $V$. \vspace{3mm}

The proof of (\ref{supero2}) uses Brownian mirror coupling techniques (cf. Section \ref{apoyy} for the basic definitions) to deal with the magnetic potential $A$. Let us mention here that the use of Brownian coupling techniques in the context of H\"older estimates for semigroups that generate diffusion (which in our case would correspond to taking $V=0$, $A=0$) has a long history, also on Riemannian manifolds (cf. \cite{Ken,Cran,LinRog} for some classical results).

\vspace{1mm}

Our main tool (cf. Theorem \ref{main} below) for the proof of (\ref{supero2}) is provided by the following estimate:\vspace{1mm}

\emph{There exists a universal constant $c_0<\infty$, such that for every $q\in (1,\infty)$, every Borel function $ A:\IR^{d}\rightarrow\IR^{d}$ with
\begin{align}\label{as2}
\max\big(\left| A\right|^{2q},  |\mathrm{div}( A)|^q\big)\in \mathcal{K}(\IR^d),
\end{align}
every $t>0$, $x\neq y$ in $\IR^d$, and every mirror coupling $(\mathsf{X},\mathsf{Y})$ of Brownian motions from $(x,y)$ one has
\begin{align*}
\IE\left(\big| e^{-\ISS_t\left( A|\mathsf{X}\right)}- e^{-\ISS_t\left( A|\mathsf{Y}\right)}\big|\right)\leq c_0 C(A,t,q)t^{-\frac{1}{2 q^*}} \left|x-y\right|^{\f{1}{q^*}},
\end{align*}
where for any Brownian motion $\mathsf{Z}$, the process $\ISS_t\left( A|\mathsf{Z}\right)$ denotes the magnetic Euclidean action functional (cf. (\ref{asasll}) below) which appears in the Feynman-Kac-It\^o formula and where the constant $C(A,t,q)<\infty$ can be computed explicitly.}\vspace{1mm}

This estimate is then combined with the Feynman-Kac-It\^o formula (and perturbation theory to deal with $V$) to finally obtain (\ref{supero2}). \vspace{2mm}

Let us mention that \emph{locally} uniform $\beta$-H\"older continuity results for nonmagnetic Schr\"odinger eigenfunctions under $L^q$-assumptions on $V$ have also been obtained in \cite{LieLos} (cf. Theorem 11.7 therein) using starightforward Sobolev embedding techniques. In addition, in \cite{Simon} (cf. Theorem B.3.5 therein) it is shown that $\beta$-dependent (Kato-type) assumptions in $V$ leading to locally uniform $\beta$-H\"older smoothing results for nonmagnetic Schr\"odinger semigroups. In the latter case, the ultimate argument relies on potential theory, while Brownian motion only enters through the Feynman-Kac formula in order to show that the Schr\"odinger semigroup is $L^{\infty}$-smoothing.  \vspace{2mm}

Using $L^p$-criteria for the $\beta$-Kato class (cf. Remark \ref{katobem}), we show that o result directly implies the following generalization of Kato's result for multi-particle Schr\"odinger operators in $\IR^{3n}$ to magnetic multi-particle Schr\"odinger operators:\vspace{1mm}

Assume there exists $\beta\in (0,1)$, $l\in\IN$ and Borel functions $a:\IR^{3}\to \IR^{3}$, $v_{i},v_{ij}:\IR^{3}\to \IR$ with 
\begin{align}\label{anmmq}
&| a|^{2/(1-\beta)}, |\mathrm{div}( a)|^{1/(1-\beta)}\in L^s(\IR^{3})+L^{\infty}(\IR^{3})\quad\text{ for some $s>3/2$},\\\label{sadd}
&v_{i}, v_{ij}\in L^s(\IR^{3})+L^{\infty}(\IR^{3})\quad\text{ for some $s>\frac{3}{2(1-\beta/2)}$  },
\end{align}
where $\mathrm{div}$ is defined in the distributional sense, and define a vector potential, resp. a magnetic potential on $\IR^{3n}$ through 
$$
A(x):=\sum^n_{i=1} a(\mathbf{x}_i),\quad V(x)=\sum_{1\leq i<j\leq n}v_{ij}(\mathbf{x}_i-\mathbf{x}_j)+\sum_{i=1}^n v_{i}(\mathbf{x}_i).
$$
Then for all $t>0$ and $p\in [1,\infty]$ one has
$$
e^{-t H (A,V)}: L^p(\IR^d)\longrightarrow C^{0,\beta}(\IR^d).
$$
To the best of our knowledge, this is the first global H\"older-regularity result for multi-particle magnetic Schr\"odinger operators.\\
Let us finally explain how this result applies to molecules in a magnetic field: Given $R\in \IR^{3n}$, $l\in\IN$, $Z\in [0,\infty)^l$, consider the potential 
$$
V_{R,Z}:\IR^{3n}\longrightarrow \IR,\quad V_{R,Z}(\mathbf{x}_1, \dots,\mathbf{x}_n):=-\sum_{i=1}^n \sum_{j=1}^l \frac{Z_j}{|\mathbf{x}_i-\mathbf{R}_j|}+ \sum_{1\leq i<j\leq n} \frac{1}{|\mathbf{x}_i-\mathbf{x}_j|}.
$$
Given $a:\IR^{3}\to \IR^{3}$ (sufficiently well-behaved), set as above $A(x):=\sum^n_{i=1} a(\mathbf{x}_i)$. Then the operator
$$
H_{A,R,Z}:=H(A,V_{R,Z})
$$
is the Hamilton operator corresponding to a molecule (in the infinite mass limit) with $l$ protons and with $n$ electrons, where the $j$-th nucleus is located in $\mathbf{R}_j$, and has $Z_j$ protons, and the electrons interact with the magnetic field induced by $A$. Then given an arbitrary $\beta\in (0,1)$ one has (\ref{sadd}) for 
$$
v_{ij}(\mathbf{x}):=1/|\mathbf{x}|,\quad v_{i}(\mathbf{x}):=-\sum^l_{j=1}\frac{Z_j}{|\mathbf{x}-\mathbf{R}_j|},
$$
so that the previous result gives that for all $t>0$ and $p\in [1,\infty]$ one has 
$$
e^{-t H(A,V)}: L^p(\IR^{3n})\longrightarrow C^{0,\beta}(\IR^{3n}),
$$
as long as
$$
| a|^{2/(1-\beta)}, |\mathrm{div}( a)|^{1/(1-\beta)}\in L^s(\IR^{3})+L^{\infty}(\IR^{3})\quad\text{ for some $s>3/2$.}
$$

\section{Main results}\label{apoyy}

We start by recalling the definition of the mirror coupling of Brownian motions as presented in \cite{HsuStu} and follow their exposition (pages 1-3 therein) closely before presenting our main results. \\
A continuous process $(\mathsf{X},\mathsf{Y})$ with values in $\IR^d\times\IR^d$ is called a \emph{coupling of Brownian motions} from $\left(x,y\right)\in\IR^d\times\IR^d$, if $\mathsf{X}$ and $\mathsf{Y}$ are Brownian motions starting in $x$ and $y$, respectively. Then, with the coupling time
$$
\tau(\mathsf{X},\mathsf{Y}):=\inf\{t>0: \mathsf{X}_s=\mathsf{Y}_s\>\text{ for all $s>t$}\},
$$
the coupling $(\mathsf{X},\mathsf{Y})$ is said to be \emph{maximal}, if for all $t>0$ one has 
\begin{align}\label{aopo}
	\IP\left(\tau(\mathsf{X},\mathsf{Y})\geq t\right)=\frac{1}{2}\int_{\IR^d} |\rho(t,x,z)-\rho(t,y,z)| dz,
\end{align}
with 
$$
(t,b)\longmapsto \rho(t,a,b)=(2\pi t)^{-d/2}e^{-\frac{|a-b|^2}{2t}}
$$
the transition density of Brownian motion starting in $a$. The reason for this notion of maximality is that for an arbitrary coupling of Brownian motions one has $\leq$ in (\ref{aopo}).\\
Let $x$ and $y$ be two distinct points of $\rr^{d}$. Then
\begin{align*}
	N_{x,y}:=\left\{v\in\rr^{d}:\left\langle v-(x+y)/2,x-y\right\rangle=0\right\},
\end{align*}
is the hyperplane orthogononal on and bisecting the segment $\underline{xy}$. Furthermore define the affine map 
\begin{align*}
R_{x,y}:\rr^{d}\longrightarrow\rr^{d},\quad	R_{x,y}v:=v-2 \left\langle v-(x+y)/2,(x-y)\left|x-y\right|^{-1}  \right\rangle(x-y)\left|x-y\right|^{-1}.
\end{align*}
This is the reflection at the hyperplane $N_{x,y}$. Let $L_{x,y}$ be the linear part of $R_{x,y}$. Note that $L_{x,y}$ is symmetric and idempotent.\\
A coupling $(\mathsf{X},\mathsf{Y})$ of Brownian motions from $(x,y)$ is called a \emph{mirror coupling}, if
\begin{align*}
	\mathsf{Y}_{t}= \begin{cases}
R_{x,y}\mathsf{X}_{t}, & t\in\left[0,\tau_{x,y}(\mathsf{X})\right], \\
\mathsf{X}_{t} & t\in\left(\tau_{x,y}(\mathsf{X}),\infty\right),
\end{cases}
\end{align*}
where 
\begin{align*}
	\tau_{x,y}(\mathsf{X}):=\inf\left\{t\geq{0}:\mathsf{X}_{t}\in{N_{x,y}} \right\}
\end{align*} 
is the hitting time of $\mathsf{X}$ with respect to $N_{x,y}$. In other words, $\mathsf{Y}$ is equal to the reflection of $\mathsf{X}$ at $N_{x,y}$ before $\mathsf{X}$ hits $N_{x,y}$, and is then equal to  $\mathsf{X}$. It follows that $\tau(\mathsf{X},\mathsf{Y})=\tau_{x,y}(\mathsf{X})$, which by an explicit calculation of $\IP(t\leq \tau_{x,y}(\mathsf{X}))$ implies that every mirror coupling is maximal.\\
Whenever well-defined, we consider the following action functional on the paths of any Brownian motion $\mathsf{Z}$, which depends on a sufficiently regular function $ A:\IR^{d}\rightarrow\IR^{d}$:
\begin{align}\label{asasll}
\ISS_t\left( A|\mathsf{Z}\right):= i  \int_{0}^{t}\langle A\left(\mathsf{Z}_{s}\right),\Id \mathsf{Z}_{s}\rangle+\frac{i}{2}\int^t_0\mathrm{div}( A)\left(\mathsf{Z}_{s}\right)\Id{s},\quad t\geq 0.
\end{align}
Above, $i$ is the imaginary unit, $\mathrm{div}( A)$ denotes the divergence of $ A$ understood in the distributional sense and the stochastic integral is understood in It\^o's sense. \\
Let $\mathbb{P}_a$ denote the law of Brownian motion starting in $a$, which is considered as a probability measure on the space of continuous paths $\omega:[0,\infty)\to \IR^d$. Generalizing the Kato class, the following hierarchy of Kato classes has been introduced in \cite{baturcd}:

\begin{Definition}\label{kat} Given $\alpha\in [0,1]$, a Borel function $f:\IR^d\to \IR$ is said to be in the \emph{$\alpha$-Kato class $\mathcal{K}^{\alpha}(\IR^d)$}, if  
$$
\lim_{t\to 0+}\sup_{z\in \IR^d}\int^t_0 s^{-\alpha/2}\int |f(\omega(s))|  \IP_z(d\omega)ds=0.
$$
\end{Definition}

Note that
\begin{align}
&\int |f(\omega(s))|  \IP_z(d\omega)= \int |f(\omega(s))|  \IP_z(d\omega)= \int_{\IR^d}\rho(t,z,y) |f(y)| dy\\
&= \int_{\IR^d}\rho(t,z,y) |f(y)| dy=(2\pi t)^{-d/2}\int_{\IR^d}e^{-\frac{|z-y|^2}{2t}} |f(y)| dy.
\end{align}

\begin{Remark}\label{katobem} 1) Each $\mathcal{K}^{\alpha}(\IR^d)$ is a linear space and $\mathcal{K}(\IR^d):=\mathcal{K}^{0}(\IR^d)$ is the usual Kato class. \\
2) One trivially has
$$
\mathcal{K}^{\alpha}(\IR^d)\subset \mathcal{K}^{\beta}(\IR^d),\quad\text{if $\alpha\geq \beta$},\\
$$
and using
$$
\int_{\IR^d}\rho(t,z,y)dy=1
$$
one gets
$$
L^{\infty}(\IR^d)\subset  \mathcal{K}^{\alpha}(\IR^d).
$$
3) For all $q\in [1,\infty)$ with $q> d/(2-\alpha)$ one has 
$$
L^q(\IR^d)+L^{\infty}(\IR^d)\subset \mathcal{K}^{\alpha}(\IR^d),
$$
which follows straightforwardly (cf. Lemma 3.9 in \cite{baturcd}) from 
$$
\rho(t,z,y)\leq C t^{-d/2}.
$$

4) For every natural $D\geq d$, every linear surjective map $\pi:\IR^D\to\IR^d$, and every $f\in  \mathcal{K}^{\alpha}(\IR^d)$ one has $f\circ\pi \in  \mathcal{K}^{\alpha}(\IR^D)$, cf. \cite{baturcd}.\\
5) For every $W\in \mathcal{K}(\IR^d)$, $z\in\IR^d$, $t>0$ one has (cf. Lemma 3.9 in \cite{baturcd})
\begin{align}\label{apqq}
\int^t_0 |W(\omega(s))|ds<\infty\quad\text{ $\mathbb{P}_z$ a.s. for all $t>0$,}
\end{align}
and if $W\in \mathcal{K}^{\beta}(\IR^d)$, then also (cf. the proof of Theorem 3.10 in \cite{baturcd})
$$
\sup_{z\in\IR^d}  \int^t_0s^{-\beta/2}\int|W(\omega(s))|  \mathbb{P}_z(d\omega)\Id{s} <\infty\quad\text{for all $t>0$.}
$$
\end{Remark}

The following probabilistic estimate is our main technical result:

\begin{Theorem}\label{main} 
There exists a universal constant $c_0<\infty$, such that for every $q\in (1,\infty)$, every Borel function $ A:\IR^{d}\rightarrow\IR^{d}$ with
\begin{align}\label{as1}
\max\big(\left| A\right|^{2q},  |\mathrm{div}( A)|^q\big)\in \mathcal{K}(\IR^d),
\end{align}
every $t>0$, $x\neq y$ in $\IR^d$, and every mirror coupling $(\mathsf{X},\mathsf{Y})$ of Brownian motions from $(x,y)$ one has
\begin{align}\label{may}
\IE\left(\big| e^{-\ISS_t\left( A|\mathsf{X}\right)}- e^{-\ISS_t\left( A|\mathsf{Y}\right)}\big|\right)\leq c_0 C(A,t,q)t^{-\frac{1}{2 q^*}} \left|x-y\right|^{\f{1}{q^*}},
\end{align}
where $1/q^*+1/q=1$ and
\begin{align*}
C(A,t,q)&:=\Big(\sup_{z\in\IR^d}  \int^t_0 \int \left| A(\omega(s))\right|^{2q}  \mathbb{P}_z(d\omega)\Id{s}\Big)^{\frac{1}{q}}\\
&\quad+\Big(\sup_{z\in\IR^d}  \int^t_0 \int\left| \frac{i}{2}\mathrm{div}(A)(\omega(s))\right|^{q}\big)  \mathbb{P}_z(d\omega)\Id{s}\Big)^{  \frac{1}{q}   }  <\infty.
\end{align*}
\end{Theorem}

\begin{Remark}1) As every Kato function is locally integrable (cf. Lemma VI.5 c) in \cite{batu2}), $\mathrm{div}(A)$ exists as a distribution in the above situation. \\
2) Remark \ref{katobem}.5 easily shows that $C(A,t,q)<\infty$ under the assumptions of Theorem \ref{main} and that $\int_{0}^{\cdot}\langle A\left(\mathsf{Z}_{s}\right),\Id \mathsf{Z}_{s}\rangle$ is a continuous $L^2$-martingale for every Brownian motion $\mathsf{Z}$ having a deterministic initial value. In particular, the process $\ISS( A|\mathsf{Z})$ is a continuous semimartingale.\\
3) The function $t\mapsto C(A,t,q)$ is locally bounded under the assumptions of Theorem \ref{main}: the easiest way to see this is to refer to Khashminiski's lemma, which implies that for every $W\in \mathcal{K}(\IR^d)$ one has 
$$
\sup_{z\in\IR^d}\int e^{\int^t_0 W(\omega(s)) ds } \IP_z(d\omega)\leq C_We^{C_Wt}\quad\text{ for all $t>0$,}
$$
and so trivially
$$
\sup_{z\in\IR^d}\int \int^t_0 W(\omega(s)) ds  \IP_z(d\omega)\leq C_We^{C_Wt}\quad\text{ for all $t>0$.}
$$

\end{Remark}

\begin{proof}[Proof of Theorem \ref{main}] Let $x\neq y$ in $\IR^{d}$ and $t>0$ be fixed. We set 
$$
\tau:=\tau(\mathsf{X},\mathsf{Y})=\tau_{x,y}(\mathsf{X}),\quad L:=L_{x,y},\quad R:=R_{x,y}.
$$
 Given a Brownian motion $\mathsf{Z}$ we split 
$$
\ISS_t(\mathsf{Z}):=\ISS_t(A|\mathsf{Z})
$$
 into
\begin{align*}
\ISS_t\left(\mathsf{Z}\right)&=\ISS^{\text{It\^o}}_t\left(\mathsf{Z}\right)+\ISS^{\text{Leb}}_t\left(\mathsf{Z}\right),\\
\ISS^{\text{It\^o}}_t\left(\mathsf{Z}\right)&:=\int_{0}^{t}{\langle A\left(\mathsf{Z}_{s}\right),\Id{\mathsf{Z}_{s}}\rangle},\\
\ISS^{\text{Leb}}\left(\mathsf{Z}\right)&:=\frac{i}{2}\int_{0}^{t}\mathrm{div}(A)\left(\mathsf{Z}_{s}\right)\Id{s}.
\end{align*}
Clearly we a.s. have
\begin{align*}
I_t&:=\int_{0}^{t}{\one_{\{s<\tau\}}\Big(\frac{i}{2}\mathrm{div}(A)\left(\mathsf{X}_{s}\right)-\frac{i}{2}\mathrm{div}(A)\left(R\mathsf{X}_{s}\right)\Big)\Id{s}}\\
&=\ISS^{\text{Leb}}_t\left(\mathsf{X}\right)-\ISS^{\text{Leb}}_t\left(\mathsf{Y}\right).
\end{align*}
Likewise, heuristically, for $s<\tau$ one has $\Id{\mathsf{Y}}_{s}=L\Id{\mathsf{X}}_{s}$ and while for $s\geq \tau$ one has $\Id{\mathsf{Y}}_{s}=\Id{\mathsf{X}}_{s}$, and we therefore expect that
\begin{align}\label{Inteq}
\ISS^{\text{It\^o}}_t\left(\mathsf{X}\right)-\ISS^{\text{It\^o}}_t\left(\mathsf{Y}\right)=M_{t}.
\end{align}
holds a.s., where
\begin{align*}
\tilde{ A}\left(x\right)&:= A\left(x\right)-L A\left(Rx\right),\\
M_{t}&:=\int_{0}^{t}{\one_{\{s<\tau\}}\langle\tilde{ A}\left(\mathsf{X}_{s}\right),\Id{\mathsf{X}}_{s}\rangle}.
\end{align*}
To show that equation (\ref{Inteq}) holds, by replacing 
$$
A=(A_1,\dots,A_d)
$$ 
with the sequence 
$$
A_n:=(\max(A_1,n),\dots, \max(A_d,n)),\quad n\in\IN, 
$$
and using the It\^o-isometry and dominated convergence, we can assume that $ A$ is bounded. By Theorem 6.5 in \cite{Ste} we have the $L^2$-convergence of the dyadic approximations
\begin{align*}
\ISS^{\text{It\^o}}_t\left(\mathsf{X}\right)-\ISS^{\text{It\^o}}_t\left(\mathsf{Y}\right)&=\lim_{n\rightarrow\infty}{\sum_{i=1}^{2^{n}-1}{\f{2^{n}}{t}\int_{t_{i-1}}^{t_{i}}{\left(\langle A\left(\mathsf{X}_{u}\right),\mathsf{X}_{t_{i+1}}-\mathsf{X}_{t_{i}}\rangle-\langle A\left(\mathsf{Y}_{u}\right),\mathsf{Y}_{t_{i+1}}-\mathsf{Y}_{t_{i}}\rangle\right)\Id{u}}}},\\
M_{t}&=\lim_{n\rightarrow\infty}{\sum_{i=1}^{2^{n}-1}{\f{2^{n}}{t}\int_{t_{i-1}}^{t_{i}}{\langle\one_{\{u<\tau\}}\tilde{ A}\left(\mathsf{X}_{u}\right),\mathsf{X}_{t_{i+1}}-\mathsf{X}_{t_{i}}\rangle\Id{u}}}},
\end{align*}
where $t_{i}:=\f{it}{2^{n}}$ for $i=0,\ldots,2^{n}$. We immediately note that in case $t<\tau$, we have $\mathsf{Y}_{s}=R\mathsf{X}_{s}$ on $\left[0,t\right]$, hence in that case by the above limits we conclude that:
\begin{align*}
\ISS^{\text{It\^o}}_t\left(\mathsf{X}\right)-\ISS^{\text{It\^o}}_t\left(\mathsf{Y}\right)=M_{t},\quad\text{for }t<\tau.
\end{align*}
If we now assume that $\tau\in\left(t_{k},t_{k+1}\right]$ for some $k=0,\ldots,2^{n}-1$, we get the following expressions for the summands in the above limits:\\
For $i\leq{k-1}$:
\begin{align*}
&\int_{t_{i-1}}^{t_{i}}{\left(\langle A\left(\mathsf{X}_{u}\right),\mathsf{X}_{t_{i+1}}-\mathsf{X}_{t_{i}}\rangle-\langle A\left(\mathsf{Y}_{u}\right),\mathsf{Y}_{t_{i+1}}-\mathsf{Y}_{t_{i}}\rangle\right)\Id{u}}\\
=&\int_{t_{i-1}}^{t_{i}}{\left(\langle A\left(\mathsf{X}_{u}\right),\mathsf{X}_{t_{i+1}}-\mathsf{X}_{t_{i}}\rangle-\langle A\left(R\mathsf{X}_{u}\right),R\mathsf{X}_{t_{i+1}}-R\mathsf{X}_{t_{i}}\rangle\right)\Id{u}}\\
=&\int_{t_{i-1}}^{t_{i}}{\langle\tilde{ A}\left(\mathsf{X}_{u}\right),\mathsf{X}_{t_{i+1}}-\mathsf{X}_{t_{i}}\rangle\Id{u}}.
\end{align*}
In the last step we have used that $L$ is selfadjoint and $Rv-Rw=L\left(v-w\right)$. \\
For $i=k$:
\begin{align*}
&\int_{t_{k-1}}^{t_{k}}{\left(\langle A\left(\mathsf{X}_{u}\right),\mathsf{X}_{t_{k+1}}-\mathsf{X}_{t_{k}}\rangle-\langle A\left(\mathsf{Y}_{u}\right),\mathsf{Y}_{t_{k+1}}-\mathsf{Y}_{t_{k}}\rangle\right)\Id{u}}\\
=&\int_{t_{k-1}}^{t_{k}}{\left(\langle A\left(\mathsf{X}_{u}\right),\mathsf{X}_{t_{k+1}}-\mathsf{X}_{t_{k}}\rangle-\langle A\left(R\mathsf{X}_{u}\right),\mathsf{X}_{t_{k+1}}-R\mathsf{X}_{t_{k}}\rangle\right)\Id{u}}.
\end{align*}
For $i=k+1$:
\begin{align*}
&\int_{t_{k}}^{t_{k+1}}{\left(\langle A\left(\mathsf{X}_{u}\right),\mathsf{X}_{t_{k+2}}-\mathsf{X}_{t_{k+1}}\rangle-\langle A\left(\mathsf{Y}_{u}\right),\mathsf{Y}_{t_{k+2}}-\mathsf{Y}_{t_{k+1}}\rangle\right)\Id{u}}\\
=&\int_{t_{k}}^{\tau}{\langle A\left(\mathsf{X}_{u}\right)- A\left(R\mathsf{X}_{u}\right),\mathsf{X}_{t_{k+2}}-\mathsf{X}_{t_{k+1}}\rangle\Id{u}},\\
&\int_{t_{k}}^{t_{k+1}}{\langle\one_{\{u<\tau\}}\tilde{ A}\left(\mathsf{X}_{u}\right),\mathsf{X}_{t_{k+2}}-\mathsf{X}_{t_{k+1}}\rangle\Id{u}}\\
=&\int_{t_{k}}^{\tau}{\langle A\left(\mathsf{X}_{u}\right)-L A\left(R\mathsf{X}_{u}\right),\mathsf{X}_{t_{k+2}}-\mathsf{X}_{t_{k+1}}\rangle\Id{u}}.
\end{align*}
For $i\geq{k+2}$ the summands vanish. Compiling these equations allows us to make the following estimates,
\begin{align*}
&\IE\left(\left|\ISS^{\text{It\^o}}_t\left(\mathsf{X}\right)-\ISS^{\text{It\^o}}_t\left(\mathsf{Y}\right)-M_{t}\right|^{2}\right)=\IE\left(\one_{\{t\geq\tau\}}\left|\ISS^{\text{It\^o}}_t\left(\mathsf{X}\right)-\ISS^{\text{It\^o}}_t\left(\mathsf{Y}\right)-M_{t}\right|^{2}\right)\\
\leq&\limsup_{n\rightarrow\infty}{\IE\left(\one_{\{t\geq\tau\}}\left|\sum_{i=1}^{2^{n}-1}{\f{2^{n}}{t}\int_{t_{i-1}}^{t_{i}}{\left(\langle A\left(\mathsf{X}_{u}\right)2-\one_{\{s<\tau\}}\tilde{ A}\left(\mathsf{X}_{u}\right),\mathsf{X}_{t_{i+1}}-\mathsf{X}_{t_{i}}\rangle\right.}}\right.\right.}\\
&\qquad\qquad\qquad\qquad\qquad\qquad\qquad{\left.\left.{{\left.-\langle A\left(\mathsf{Y}_{u}\right),\mathsf{Y}_{t_{i+1}}-\mathsf{Y}_{t_{i}}\rangle\right)\Id{u}}}\right|^{2}\right)}\\
=&\limsup_{n\rightarrow\infty}{\sum_{k=0}^{2^{n}-1}{\IE\left(\one_{\{\tau\in\left(t_{k},t_{k+1}\right]\}}\left|\sum_{i=1}^{2^{n}-1}{\f{2^{n}}{t}\int_{t_{i-1}}^{t_{i}}{\left(\langle A\left(\mathsf{X}_{u}\right)-\one_{\{s<\tau\}}\tilde{ A}\left(\mathsf{X}_{u}\right),\mathsf{X}_{t_{i+1}}-\mathsf{X}_{t_{i}}\rangle\right.}}\right.\right.}}\\
&\qquad\qquad\qquad\qquad\qquad\qquad\qquad\qquad\qquad\ {{\left.\left.{{\left.-\langle A\left(\mathsf{Y}_{u}\right),\mathsf{Y}_{t_{i+1}}-\mathsf{Y}_{t_{i}}\rangle\right)\Id{u}}}\right|^{2}\right)}}\\
=&\limsup_{n\rightarrow\infty}{\sum_{k=0}^{2^{n}-1}{\IE\left(\one_{\{\tau\in\left(t_{k},t_{k+1}\right]\}}\left|\f{2^{n}}{t}\left(\int_{t_{k-1}}^{t_{k}}{\langle A\left(R\mathsf{X}_{u}\right),\left(R-\id\right)\mathsf{X}_{t_{k+1}}\rangle\Id{u}}\right.\right.\right.}}\\
&\qquad\qquad\qquad\qquad\qquad\qquad\qquad{{\left.\left.\left.+\int_{t_{k}}^{\tau}{\langle\gamma\left(\mathsf{X}_{u}\right),\mathsf{X}_{t_{k+2}}-\mathsf{X}_{t_{k+1}}\rangle\Id{u}}\right)\right|^{2}\right)}},
\end{align*}
where $\gamma\left(z\right):=L A\left(Rz\right)- A\left(Rz\right)$. Note that 
$$
\left(R-\id\right)\mathsf{X}_{t_{k+1}}=L\left(\mathsf{X}_{t_{k+1}}-\mathsf{X}_{\tau}\right)-\left(\mathsf{X}_{t_{k+1}}-\mathsf{X}_{\tau}\right),
$$
because $R\mathsf{X}_{\tau}=\mathsf{X}_{\tau}$. In particular, since $L$ is self-adjoint and idempotent,
$$
\left|\left(R-\id\right)\mathsf{X}_{t_{k+1}}\right|^2=\left|\mathsf{X}_{t_{k+1}}-\mathsf{X}_{\tau}\right|^2.
$$
Since $ A$ is bounded by some $\kappa>0$, and so $|\gamma|\leq 2\kappa$, using 
$$
\left(a+b\right)^2\leq 2a^2+2b^2, \quad t_{j+1}-t_j=\frac{t}{2^n}, 
$$
we can thus estimate as follows,
\begin{align*}
&\IE\left(\left|\ISS^{\text{It\^o}}_t\left(\mathsf{X}\right)-\ISS^{\text{It\^o}}_t\left(\mathsf{Y}\right)-M_{t}\right|^{2}\right)\\
\leq&\limsup_{n\rightarrow\infty}\sum_{k=0}^{2^{n}-1}\f{2^{n+2}}{t^2}\IE\left(\one_{\{\tau\in\left(t_{k},t_{k+1}\right]\}}\left(\int_{t_{k-1}}^{t_{k}}{\left|A\left(R\mathsf{X}_{u}\right) \right|\left|\left(R-\id\right)\mathsf{X}_{t_{k+1}}\right|\Id{u}}\right)^2\right)\\
&\quad+ \limsup_{n\rightarrow\infty}\sum_{k=0}^{2^{n}-1}\f{2^{n+2}}{t^2}\IE\left(\one_{\{\tau\in\left(t_{k},t_{k+1}\right]\}}\left(\int_{t_{k}}^{\tau}{\left|\gamma\left(\mathsf{X}_{u}\right)\right|\left|\mathsf{X}_{t_{k+2}}-\mathsf{X}_{t_{k+1}}\right|\Id{u}}\right)^{2}\right)\\
\leq&\limsup_{n\rightarrow\infty}4\kappa^2\sum_{k=0}^{2^{n}-1}\IE\left(\one_{\{\tau\in\left(t_{k},t_{k+1}\right]\}}\left|\left(R-\id\right)\mathsf{X}_{t_{k+1}}\right|^2\right)\\
&\quad+ 16\kappa^2\limsup_{n\rightarrow\infty}\sum_{k=0}^{2^{n}-1}\IE\left(\one_{\{\tau\in\left(t_{k},t_{k+1}\right]\}}\left|\mathsf{X}_{t_{k+2}}-\mathsf{X}_{t_{k+1}}\right|^{2}\right)\\
\leq&16\kappa^2\limsup_{n\rightarrow\infty}{\sum_{k=0}^{2^{n}-1}{\left(\IE\left(\one_{\{\tau\in\left(t_{k},t_{k+1}\right]\}}\left|\mathsf{X}_{t_{k+2}}-\mathsf{X}_{t_{k+1}}\right|^{2}\right)+\IE\left(\one_{\{\tau\in\left(t_{k},t_{k+1}\right]\}}\left|\mathsf{X}_{t_{k+1}}-\mathsf{X}_{\tau}\right|^{2}\right)\right)}}.
\end{align*}
Since $\tau$ is an $\mathsf{X}$-stopping time, we conclude by the Markov property of $\mathsf{X}$,
using
$$
\int |\omega(r)-\omega(s)|^2 \IP_z(d\omega)= r-s, \quad r>s>0,
$$
and once more $t_{j+1}-t_j=\frac{t}{2^n}$, that
\begin{align*}
&\IE\left(\left|\ISS^{\text{It\^o}}_t\left(\mathsf{X}\right)-\ISS^{\text{It\^o}}_t\left(\mathsf{Y}\right)-M_{t}\right|^{2}\right)\\
\leq&16\kappa^2\limsup_{n\rightarrow\infty}{\sum_{k=0}^{2^{n}-1}{\left(\f{t}{2^{n}}\IP\left(\tau\in\left(t_{k},t_{k+1}\right]\right)+\int_{t_{k}}^{t_{k+1}}{\IE\left(\left.\left|\mathsf{X}_{t_{k+1}}-\mathsf{X}_{u}\right|^{2}\right|\tau=u\right)\tau_*\IP\left(du\right)}\right)}}\\
\leq&16\kappa^2\limsup_{n\rightarrow\infty}{\f{t}{2^{n}}\IP\left(t\geq\tau\right)+16\kappa^2\limsup_{n\rightarrow\infty}\f{t}{2^{n}}\sum_{k=0}^{2^{n}-1}{\int_{t_{k}}^{t_{k+1}}{\tau_*\IP\left(du\right)}}}\\
=&\limsup_{n\rightarrow\infty}{\f{32\kappa^2{t}}{2^{n}}\IP\left(t\geq\tau\right)}=0.
\end{align*}
Altogether, we have found that under the assumptions of the theorem one has
\begin{align}\label{nase}
\ISS_t\left(\mathsf{X}\right)-\ISS_t\left(\mathsf{Y}\right)=M_{t}+I_t\quad\text{a.s.}
\end{align}
We are now going to estimate the $L^1$-norms of $M_ {t}$ and $I_t$. Let us start with $\IE\left(\left|I_t\right|\right)$: setting
$$
w:=\frac{i}{2}\mathrm{div}(A),\quad p:=q^*,
$$
we have
\begin{align*}
\IE\left(\left|I_t\right|\right)&\leq\int_{0}^{t}{\IE\left(\left|\one_{\{s<\tau\}}\left(w\left(\mathsf{X}_{s}\right)-w\left(R\mathsf{X}_{s}\right)\right)\right|\right)\Id{s}}\\
&\leq\left(\int_{0}^{t}{\IP\left(s<\tau\right)\Id{s}}\right)^{\f{1}{p}} \left(\int_{0}^{t}\IE\left(\left|w\left(\mathsf{X}_{s}\right)-w\left(R\mathsf{X}_{s}\right)\right|^{q}\right)\Id{s}\right)^{\f{1}{q}}   \\
&\leq 2\left(\int_{0}^{t}{\IP\left(s<\tau\right)\Id{s}}\right)^{\f{1}{p}}C_{w,t,q},
\end{align*}
where
$$
C_{w,t,q}:=\left(\sup_{z\in\IR^{d}}\int_{0}^{t}{\int\left|w\left(\omega(s)\right)\right|^{q}\mathbb{P}_z(d\omega)\Id{s}}\right)^{\f{1}{q}}.
$$
In view of 
\begin{align*}
	\IP\left(\tau>s\right)&\leq\frac{1}{2}\int_{\IR^d} |\rho(s,x,z)-\rho(s,y,z)| dz=\frac{2}{\sqrt{2\pi s}}\int_{0}^{\frac{|x-y|}{2}}{e^{-\frac{u^{2}}{2s}}\Id u}\\
	&\leq \left(2\pi\right)^{-1/2}|x-y|s^{-1/2}  ,
	\end{align*}
	we conclude
\begin{align*}
\IE\left(\left|I_t\right|\right)&\leq CC_{w,t,q}t^{-\frac{1}{2p}}|x-y|^{\frac{1}{p}},
\end{align*}
where from here on $C<\infty$ denotes a universal constant whose value may change from line to line. Now let us turn to the estimate for $\IE\left(\left|M_{t}\right|\right)$: define
\begin{align*}
h\left(r\right):=\begin{cases} \f{1}{12}\left(2-\left|r\right|\right)^{3}+\left|r\right|, & \left|r\right|\leq{2},\\
								\left|r\right|, & \left|r\right|>2.
				\end{cases}
\end{align*}
We note that $h\left(r\right)\geq\left|r\right|$ and that $h$ is in $C^{2}\left(\IR\right)$ with
\begin{align*}
h''\left(r\right)=\begin{cases} 1-\f{1}{2}\left|r\right|, & \left|r\right|\leq{2},\\
								0, & \left|r\right|>2.
				\end{cases}
\end{align*}
In particular we have $|h''|\leq\one_{\left[-2,2\right]}$ and $\left|h'\right|\leq{1}$. We conclude by It\^o's formula
\begin{align*}
h\left(M_{t}\right)=\f{1}{2}\int_{0}^{t}{h''\left(M_{s}\right)\one_{\{s<\tau\}}\left|\tilde{ A}\left(\mathsf{X}_{s}\right)\right|^{2}\Id{s}}+\tilde{M}_{t},
\end{align*}
where 
\begin{align*}
\tilde{M}_{t}=\int_{0}^{t}{h'\left(M_{s}\right)\one_{\{s<\tau\}}\langle\tilde{ A}\left(\mathsf{X}_{s}\right),\Id{\mathsf{X}_{s}}\rangle}.
\end{align*}
is an ($L^2$)-martingale, as follows from the assumption on $ A$ and the boundedness of $h'$. Thus 
$$
\IE (\tilde{M}_{t} )=\IE (\tilde{M}_{0} )=0,
$$
and we have
\begin{align*}
\IE\left(\left|M_{t}\right|\right)&\leq\IE\left(h\left(M_{t}\right)\right)\\
&=\f{1}{2}\IE\left(\int_{0}^{t}{h''\left(M_{s}\right)\one_{\{s<\tau\}}\left|\tilde{ A}\left(\mathsf{X}_{s}\right)\right|^{2}\Id{s}}\right)\\
&\leq\f{1}{2}\IE\left(\int_{0}^{t}{\one_{\{s<\tau\}}\left|\tilde{ A}\left(\mathsf{X}_{s}\right)\right|^{2}\Id{s}}\right)\\
&\leq\f{1}{2}\left(\int_{0}^{t}{\IP\left(s<\tau\right)\Id{s}}\right)^{\f{1}{p}} \left(\int_{0}^{t}{\IE\left(\left|\tilde{ A}\left(\mathsf{X}_{s}\right)\right|^{2q}\right)\Id{s}}\right)^{\f{1}{q}}\\
&\leq\left(\int_{0}^{t}{\IP\left(s<\tau\right)\Id{s}}\right)^{\f{1}{p}}{\left(\sup_{z\in\IR^{d}}\int_{0}^{t}{\int\left| A\left(\omega(s)\right)\right|^{2q}\mathbb{P}_z(\omega)\Id{s}}\right)^{\f{1}{q}}}.
\end{align*}
Hence, similarly to the Lebesgue integral $I_t$, we conclude 
\begin{align*}
\IE\left(\left|M_{t}\right|\right)\leq C C_{A,t,q}t^{-\frac{1}{2p}}\left|x-y\right|^{\f{1}{p}},
\end{align*}
where
$$
C_{A,t,q}:=\left(\sup_{z\in\IR^{d}}\int_{0}^{t}\int \left| A\left(\omega(s)\right)\right|^{2q}\IP_z(d\omega)\Id s\right)^{\f{1}{q}}.
$$
Thus we have shown
\begin{align*}
\IE\left(\left|\ISS_t\left(\mathsf{X}\right)-\ISS_t\left(\mathsf{Y}\right)\right|\right)\leq C(C_{w,t,q}+C_{A,t,q})t^{-\frac{1}{2p}}\left|x-y\right|^{\f{1}{p}}.
\end{align*}
Finally, noting that for all purely imaginary $z,z'$ one has the elementary estimate
\begin{align*}
\left|e^{z}-e^{z'}\right|\leq C \left|z-z'\right|,
\end{align*}
and $\Re(\ISS_t\left(\mathsf{Z}\right))=0$, the proof is complete.
\end{proof}

If $ A:\IR^{d}\rightarrow\IR^{d}$ and $V:\IR^{d}\rightarrow\IR$ are Borel functions with 
\begin{align}\label{hops}
\max\big(\left|  A\right|^{2},  | \mathrm{div}( A)|,\left|V\right|\big)\in \mathcal{K}(\IR^d),
\end{align}
then the symmetric sesquilinear form 
$$
 (\Psi_1,\Psi_2)\longmapsto \frac{1}{2}\int_{\IR^d} \big((i\nabla+A) \Psi_1,(i\nabla+A)\nabla\Psi_2\big) + \int_{\IR^d}V\cdot\overline{\Psi_1}\cdot\Psi_2 \in \IC
$$
in $L^2(\IR^d)$ with domain of definition $C^{\infty}_c(\IR^d)$ is semibounded from below and closable \cite{BroHunLes}. Thus, the closure of this form induces a self-adjoint semibounded from below operator $H(A,V)$ in $L^2(\IR^d)$. The corresponding magnetic Schr\"odinger semigroup is given by the Feynman-Kac-It\^o formula \cite{BroHunLes} 
$$
e^{-tH(A,V)} \Psi(x)=\IE\left(e^{-\ISS_t(A|\mathsf{Z}(x))-\int^t_0V(\mathsf{Z})ds}\Psi(\mathsf{Z}_t(x))\right), \quad \Psi\in L^2(\IR^d),
$$
where $\mathsf{Z}(x)$ is an arbitrary Brownian motion in $\IR^d$ starting in $x\in \IR^d$. Using the Feynman-Kac-It\^o formula for $V=0$ with Theorem \ref{main} to deal with the magnetic potential $A$, and perturbation theory to deal with the electric potential $V$, we can now establish:

\begin{Theorem}\label{main2} Let $\beta \in (0,1)$, let $ A:\IR^{d}\rightarrow\IR^{d}$, $V:\IR^d\to\IR$ be Borel functions which satisfy 
\begin{align}\label{super}
\max\big(\left| A\right|^{\frac{2}{1-\beta}},  |\mathrm{div}( A)|^{\frac{1}{1-\beta}}\big)\in\mathcal{K}(\IR^d),\quad V\in\mathcal{K}^{\beta}(\IR^d),
\end{align}
and let $t>0$, $1\leq p\leq q\leq \infty$. Then one has
$$
e^{-t H (A,V)}: L^p(\IR^d)\longrightarrow C^{0,\beta}(\IR^d)\cap L^q(\IR^d)
$$  
continuously, and there exists a universal constant $c_0<\infty$ and a constant $C_V<\infty$ which only depends on $V$, such that
\begin{align*}
&\left\|e^{-tH(A,V)}\right\|_{L^{p}\to C^{0,\beta}\cap L^q}\\
&\leq C_V t^{-\frac{d}{2}(\frac{1}{p}-\frac{1}{q})}e^{C_Vt}+\left(c_0C(A,t/2,(1-\beta)^{-1})\left(\frac{t}{2}\right)^{-\frac{\beta}{2 }} + c_0\left(\frac{t}{2}\right)^{-\frac{\beta}{2}} \right.\\
&\left.\quad+ C_{V}e^{C_V\frac{t}{2}}\int^{t/2}_0\big( C(A,s/2,(1-\beta)^{-1})s^{-\frac{\beta}{2 }} + s^{-\frac{\beta}{2}}\big) D_V(s/2)ds\right)C_V \left(\frac{t}{2}\right)^{-\frac{d}{2}(\frac{1}{p})}e^{C_V\frac{t}{2}},
\end{align*}
where 
$$
(0,\infty)\ni s\longrightarrow C(A,s,(1-\beta)^{-1})\in[0,\infty)
$$
is the locally bounded function from Theorem \ref{main}, and
$$
D_V:(0,\infty)\longrightarrow [0,\infty],\quad D_V(s):=\sup_{z\in\IR^d}\int |V(\omega(s))|    \IP_z(d\omega).
$$
\end{Theorem}

\begin{Remark}1) Using monotone convergence one finds
$$
\int^t_0 s^{-\beta/2}\sup_{z\in\IR^d}\int |V(\omega(s))|    \IP_z(d\omega)ds\leq \sup_{z\in\IR^d}\int^t_0s^{-\beta/2}\int |V(\omega(s))|    \IP_z(d\omega)ds, 
$$
which is finite for all $t>0$ by Remark \ref{katobem}.5), so that a posteriori one also has $D_V<\infty$ a.e.\\
2) As our proof shows, the constant $C_V$ can be chosen to be any constant which satisfies that for all $1\leq p\leq q\leq \infty$, $r>0$ one has 
 \begin{align*}
\left\|e^{-rH(A,V)}\right\|_{L^p\to L^q}\leq C_V r^{-\frac{d}{2}(\frac{1}{p}-\frac{1}{q})}e^{C_Vr}.
\end{align*}
The existence of such a uniform constant has been shown in \cite{BroHunLes}.\\
3) Using $\Psi=e^{t \IR^d}e^{-t H(A,V)}\Psi$ for eigenfunctions $\Psi$ of $H(A,V)$, one obtains explicit \mbox{$L^r\to C^{0,\beta}$}-estimates for eigenfunctions.
\end{Remark}

\begin{proof}[Proof of Theorem \ref{main2}] We start by noting that the assumptions on $A$ together with Jensen's inequality, and that $\mathcal{K}^{\beta}(\IR^d)\subset \mathcal{K}(\IR^d)$ shows that the pair $(A,V)$ satisfies (\ref{hops}).\\
Set $q:=1/(1-\beta)\in (1,\infty)$ so that $q^*=1/\beta$ and pick a mirror coupling $(\mathsf{X},\mathsf{Y})$ from $(x,y)\in (\IR^d\times \IR^d)\setminus \mathrm{diag}(\IR^d)$ and set $\tau:=\tau(\mathsf{X},\mathsf{Y})$. Then, given $r>0$, $\Phi\in L^2(\IR^d)\cap L^{\infty}(\IR^d)$ we can estimate as follows,
\begin{align*}
& \left|e^{-rH(A,0)}\Phi(x)-e^{-rH(A,0)}\Phi(y)\right|\\
&\leq\IE\left(\left|e^{-\ISS_{r}\left(A|\mathsf{X}\right)}-e^{-\ISS_{r}\left(A|\mathsf{Y}\right)}\right||\Phi(\mathsf{X}_r)|\right)+\IE\left(\left|e^{-\ISS_{r}\left(A|\mathsf{X}\right)}\right|\left|\Phi(\mathsf{X}_r)-\Phi(\mathsf{Y}_r)\right|\right)\\
&=\IE\left(\left|e^{-\ISS_{r}\left(A|\mathsf{X}\right)}-e^{-\ISS_{r}\left(A|\mathsf{Y}\right)}\right||\Phi(\mathsf{X}_r)|\right)+\IE\left(1_{\{r<\tau\}}\left|e^{-\ISS_{r}\left(A|\mathsf{X}\right)}\right|\left|\Phi(\mathsf{X}_r)-\Phi(\mathsf{Y}_r)\right|\right)\\
&\leq \left\|\Phi\right\|_{\infty} \IE\left(\left|e^{-\ISS_{r}\left(A|\mathsf{X}\right)}-e^{-\ISS_{r}\left(A|\mathsf{Y}\right)}\right|\right)+\IE\left(1_{\{r<\tau\}}\left|\Phi(\mathsf{X}_r)-\Phi(\mathsf{Y}_r)\right|\right)\\
&\leq C(A,r,q)r^{-\frac{1}{2 q^*}} \left|x-y\right|^{\f{1}{q^*}} \left\|\Phi\right\|_{\infty}+ 2\IP(r<\tau)\left\|\Phi\right\|_{\infty}\\
&\leq C(A,r,q)r^{-\frac{1}{2 q^*}} \left|x-y\right|^{\f{1}{q^*}} \left\|\Phi\right\|_{\infty}+ 2\IP(r<\tau)^{1/q^*}\left\|\Phi\right\|_{\infty}\\
&\leq C(A,r,q)r^{-\frac{1}{2 q^*}} \left|x-y\right|^{\f{1}{q^*}} \left\|\Phi\right\|_{\infty}+ c_0r^{-\frac{1}{2q^*}}\left|x-y\right|^{\f{1}{q^*}} \left\|\Phi\right\|_{\infty},
\end{align*}
where $c_0<\infty$ is a universal constant. Thus we have shown
$$
\left\|e^{-rH(A,0)}\right\|_{L^{\infty}\to C^{0,\beta}}\leq c_0C(A,r,1/(1-\beta))r^{-\frac{\beta}{2 }} + c_0r^{-\frac{\beta}{2}}.
$$
Duhamel's formula\footnote{In principle one should be more careful here as $V$ is not bounded; but for the purpose of proving the estimate from Theorem \ref{main2} one can approximate $V$ with a sequence of bounded potentials $V_n$ and take $n\to\infty$ in the end (cf. the proof Theorem 3.10 in \cite{baturcd}).} states that
\begin{align*}
&e^{-tH(A,V)}\Phi=e^{-tH(A,0)}\Phi+\int^{t}_0 e^{-\frac{s}{2}H(A,0)}e^{-\frac{s}{2}H(A,0)}Ve^{-(t-s)H(A,V)}\Phi ds,
\end{align*}
and so
\begin{align}\label{v4}
&\left\|e^{-tH(A,V)}\Phi\right\|_{C^{0,\beta}}\leq\left\|e^{-tH(A,0)}\Phi\right\|_{C^{0,\beta}}\\\nn
&\>\>\quad+\int^{t}_0 \left\|e^{-\frac{s}{2}H(A,0)}\right\|_{L^{\infty}\to C^{0,\beta}}\left\|e^{-\frac{s}{2}H(A,0)}V\right\|_{L^{\infty}\to L^{\infty}}\left\|e^{-(t-s)H(A,V)}\Phi \right\|_{L^{\infty}}ds.
\end{align}
There exists \cite{BroHunLes} a constant $C_V$ such that for all $1\leq p\leq q\leq \infty$, $r>0$ one has 
 \begin{align}\label{glat}
\left\|e^{-rH(A,V)}\right\|_{L^p\to L^q}\leq C_V r^{-\frac{d}{2}(\frac{1}{p}-\frac{1}{q})}e^{C_Vr},
\end{align}
so that
\begin{align}\label{v1}
\left\|e^{-(t-s)H(A,V)}\Phi\right\|_{\infty}\leq C_V e^{C_Vt}\left\|\Phi\right\|_{\infty}.
\end{align}
Moreover, by what we have shown above, 
\begin{align}\label{v3}
\left\|e^{-\frac{s}{2}H(A,0)}\right\|_{L^{\infty}\to C^{0,\beta}}\leq 
CC(A,s/2,1/(1-\beta))s^{-\frac{\beta}{2 }} + Cs^{-\frac{\beta}{2}}.
\end{align}
Given $f\in L^{\infty}(\IR^d)$, $x\in \IR^d$, and a Brownian motion $\mathsf{Z}(x)$ in $\IR^d$ starting in $x$ we have, using $|e^{-\ISS_{s/2}(A|\mathsf{Z}(x)) } |=1$, the estimate
\begin{align*}
&\left|e^{-\frac{s}{2}H(A,0)}Vf(x)\right|=\left|\IE\left(e^{-\ISS_{s/2}(A|\mathsf{Z}(x)) }   V(\mathsf{Z}_{s/2}(x))f(\mathsf{Z}_{s/2}(x))\right)\right|\\
\leq&\IE\left(V(\mathsf{Z}_{s/2}(x))f(\mathsf{Z}_{s/2}(x))\right)=\int|V(\omega(s/2)) |\cdot |f(\omega(s/2))| \IP_x(d\omega)\\
\leq&\left\|f\right\|_{\infty} D_V(s/2),\\
\end{align*}
so that
\begin{align}\label{v2}
\left\|e^{-\frac{s}{2}H(A,0)}V\right\|_{L^{\infty}\to L^{\infty}}\leq D_V(s/2).
\end{align}
Combining (\ref{v4}), (\ref{v1}), (\ref{v3}), (\ref{v2}) we have shown that for all $\Phi\in L^{\infty}(\IR^d)$,
\begin{align*}
&\left|e^{-\frac{t}{2}H(A,V)}\Phi(x)-e^{-\frac{t}{2}H(A,V)}\Phi(y)\right|\\
&\leq\left(c_0C(A,t/2,(1-\beta)^{-1})\left(\frac{t}{2}\right)^{-\frac{\beta}{2 }} + c_0\left(\frac{t}{2}\right)^{-\frac{\beta}{2}}\right.\\
&\quad\left.+ C_{V}e^{C_V\frac{t}{2}}\int^{t/2}_0\big( C(A,s/2,(1-\beta)^{-1})s^{-\frac{\beta}{2 }} + s^{-\frac{\beta}{2}}\big) D_V(s/2)ds \right) \left\|\Phi\right\|_{L^{\infty}}|x-y|^{\beta},
\end{align*}
and so
\begin{align*}
&\left\|e^{-\frac{t}{2}H(A,V)}\right\|_{L^{\infty}\to C^{0,\beta}}\\
& \leq c_0C(A,t/2,(1-\beta)^{-1})\left(\frac{t}{2}\right)^{-\frac{\beta}{2 }} + c_0\left(\frac{t}{2}\right)^{-\frac{\beta}{2}} \\
&\quad+ C_{V}e^{C_V\frac{t}{2}}\int^{t/2}_0\big( C(A,s/2,(1-\beta)^{-1})s^{-\frac{\beta}{2 }} + s^{-\frac{\beta}{2}}\big) D_V(s/2)ds
\end{align*}
The above estimate together with $\Phi=e^{-\frac{t}{2}H(A,V)}\Psi$ and (\ref{glat}) shows 
\begin{align*}
&\left\|e^{-tH(A,V)}\right\|_{L^{p}\to C^{0,\beta}}\\
&\leq \left\|e^{-\frac{t}{2}H(A,V)}\right\|_{L^{\infty}\to C^{0,\beta}}\left\|e^{-\frac{t}{2}H(A,V)}\right\|_{L^{p}\to L^{\infty}}\\
&\leq \left(c_0C(A,t/2,(1-\beta)^{-1})\left(\frac{t}{2}\right)^{-\frac{\beta}{2 }} + c_0\left(\frac{t}{2}\right)^{-\frac{\beta}{2}} \right.\\
&\left.\quad+ C_{V}e^{C_V\frac{t}{2}}\int^{t/2}_0\big( C(A,s/2,(1-\beta)^{-1})s^{-\frac{\beta}{2 }} + s^{-\frac{\beta}{2}}\big) D_V(s/2)ds\right)C_V \left(\frac{t}{2}\right)^{-\frac{d}{2}(\frac{1}{p})}e^{C_V\frac{t}{2}}.
\end{align*}
Finally, using (\ref{glat}) we end up with
\begin{align*}
&\left\|e^{-tH(A,V)}\right\|_{L^{p}\to C^{0,\beta}\cap L^q}\\
&\leq C_V t^{-\frac{d}{2}(\frac{1}{p}-\frac{1}{q})}e^{C_Vt}+\left(c_0C(A,t/2,(1-\beta)^{-1})\left(\frac{t}{2}\right)^{-\frac{\beta}{2 }} + c_0\left(\frac{t}{2}\right)^{-\frac{\beta}{2}} \right.\\
&\left.\quad+ C_{V}e^{C_V\frac{t}{2}}\int^{t/2}_0\big( C(A,s/2,(1-\beta)^{-1})s^{-\frac{\beta}{2 }} + s^{-\frac{\beta}{2}}\big) D_V(s/2)ds\right)C_V \left(\frac{t}{2}\right)^{-\frac{d}{2}(\frac{1}{p})}e^{C_V\frac{t}{2}},
\end{align*}
which completes the proof.
\end{proof}

For the following result consider the linear surjective maps
\begin{align*}
&\pi_j:\IR^{3n}\longrightarrow \IR^3, \quad (\mathbf{x}_1,\dots,\mathbf{x}_n)\longmapsto \mathbf{x}_j,\\
&\pi_{ij}:=\pi_i-\pi_j:\IR^{3n}\longrightarrow \IR^3,
\end{align*}
and let $A:\IR^{3n}\to \IR^{3n}$, $V:\IR^{3n}\to\IR$ be arbitrary functions. Remark \ref{katobem} then shows:

\begin{Corollary}\label{poss} Let $\beta\in (0,1)$, $l\in\IN$ and let $a:\IR^{3}\to \IR^{3}$, $v_{i},v_{ij}:\IR^{3}\to \IR$ be Borel functions with 
$$
A=\sum^n_{i=1} a\circ \pi_i,\quad V=\sum_{1\leq i<j\leq n}v_{ij}\circ\pi_{ij}+\sum_{i=1}^n v_{i}\circ\pi_{i},
$$
and
\begin{align*}
&| a|^{2/(1-\beta)}, |\mathrm{div}( a)|^{1/(1-\beta)}\in L^s(\IR^{3})+L^{\infty}(\IR^{3})\quad\text{ for some $s>3/2$},\\
&v_{i}, v_{ij}\in L^s(\IR^{3})+L^{\infty}(\IR^{3})\quad\text{ for some $s>\frac{3}{2(1-\beta/2)}$  }.
\end{align*}
Then for all $t>0$ and $p\in [1,\infty]$ one has
$$
e^{-t H (A,V)}: L^p(\IR^{3n})\longrightarrow C^{0,\beta}(\IR^{3n})
$$
continuously.
\end{Corollary}

%
%
%
%
%


\end{document}